\documentstyle[prd,aps,floats]{revtex}

\begin{document}

\draft

\input epsf
\renewcommand{\topfraction}{0.8}
\renewcommand{\bottomfraction}{0.8}
\twocolumn[\hsize\textwidth\columnwidth\hsize\csname 
@twocolumnfalse\endcsname

\title{Dilatonic Domain Walls and 
Curved Intersecting Branes}
\author{James E. Lidsey} 
\address{Astronomy Unit, School of Mathematical Sciences,
Queen Mary, University of London, Mile End Road, \\
London, E1 4NS,~~U.~~K. \\
J.E.Lidsey@qmw.ac.uk}
\maketitle
\begin{abstract}
Curved, intersecting brane configurations
satisfying the 
type IIA supergravity
equations of motion are found. 
In eleven dimensions, the models are interpreted 
in terms of 
orthogonally intersecting M5--branes, 
where 
the world--volumes are curved due to 
the effects of one or more massless 
scalar fields. Duality symmetries are 
employed to generate further 
type II and heterotic solutions. 
Some cosmological implications are discussed. 

\end{abstract}


\vskip2pc]

Intersecting brane configurations 
in supergravity 
theories have played a
fundamental  
role 
in uncovering the conjectured 
duality symmetries that relate the five  
perturbative string theories. (For a review, see, e.g., Refs. 
\cite{argurio,gauntlett}).  In this paper,  
we find a wide class of 
(non--supersymmetric) intersecting, 
curved branes within the context 
of the $D=10$, $N=2$ and $D=11$, $N=1$
supergravity theories. These solutions provide 
the seeds for generating further configurations through 
duality transformations. Moreover, since the
world--volumes of the branes  
are curved, they admit a cosmological interpretation and 
can therefore provide a framework for addressing 
questions arising in 
the recently proposed braneworld scenario 
\cite{braneref,hw,lukas,RS}. 
Intersecting 
branes with curved world--volumes were 
recently derived from string dualities 
in Ref. \cite{prt}.

The sector of a $D$--dimensional supergravity action 
that leads to a solitonic (magnetically charged) $p$--brane
is given by 
\footnote{In this paper, 
spacetime has signature $(-, + , \ldots , +)$. Variables in ten
and eleven dimensions are represented with 
the accents $\hat{}$ and $\check{}$, respectively. 
Upper case, Latin indices take values in the 
range $A=(0, 1, \ldots , D-1 )$. 
Lower case Greek (Latin) indices correspond to world--volume (transverse 
space) coordinates. A totally antisymmetric $p$--form 
is defined by $A_{(p)} = (1/p!) A_{A_1 \ldots A_p} 
dx^{A_1}\wedge \ldots \wedge dx^{A_p}$ 
and has a 
field strength given by the exterior 
derivative $F_{(p+1)} = dA_{(p)}$.
The 
Chern--Simons terms arising in the bosonic sectors of the theories
are trivial 
for the configurations we consider.} 
\begin{equation}
\label{effectiveaction}
S=\int d^D x \sqrt{|g|} \left[ R -\frac{1}{2} \left( \nabla 
\Phi \right)^2 -\frac{1}{2q!}e^{\alpha \Phi} F^2_{(q)} \right]  ,
\end{equation}
where 
$R$ is the Ricci curvature scalar of the spacetime with metric, 
$g_{AB}$, $g \equiv {\rm det}g_{AB}$ 
and 
the constant, $\alpha$, 
parametrizes the coupling between the form 
field and the $D$--dimensional dilaton, $\Phi$. 

A domain wall in $D$ dimensions may be viewed 
as a solitonic $(D-2)$--brane 
supported by a $0$--form field strength, i.e., 
a cosmological constant, 
$F_{(0)}^2 = \Lambda^2$.
The metric and dilaton field are given by
\begin{eqnarray}
\label{isotropicbrane1}
ds^2_D = H^m f_{\mu\nu} dx^{\mu}dx^{\nu} + H^n 
dy^2 \\
\label{isotropicbrane2}
e^{\Phi} = H^{-2\alpha / \Delta}   ,
\end{eqnarray}
where
\begin{eqnarray}
\label{mnDelta}
m \equiv \frac{4}{\Delta (D-2)} , \qquad n \equiv 
\frac{4 (D-1)}{\Delta (D-2)} \nonumber \\
\Delta \equiv \alpha^2 -\frac{2(D-1)}{D-2} 
\end{eqnarray}
are constants \cite{lpss}.
The harmonic function is given by 
$H=1 +m |y|$, 
where $m^2 = \Delta \Lambda^2 /4$ and 
$y$ is the coordinate on the transverse dimension. 
The mass parameter, $m$, should be viewed as 
a
piecewise constant of integration arising from a 
dual reformulation of action (\ref{effectiveaction}) 
in terms of a $D$--form field strength.
The world--volume metric,
$f_{\mu\nu}$, is  
the metric on a $(D-1)$--dimensional 
Ricci--flat spacetime \cite{brecher}.

In the present analysis, we consider the 
more general metric ansatz \cite{lidsey,os,feinstein}: 
\begin{equation}
\label{curvedmetric}
ds^2_D =H^m f_{\mu\nu} dx^{\mu}dx^{\nu} +e^{2B} H^n 
dy^2   ,
\end{equation}
where $\{ m,n  ,H(y) \}$ are defined 
as above 
and the function $B=B(x)$ depends only on the world--volume coordinates.
It can then be verified by direct substitution 
that the field equations derived by varying the action 
(\ref{effectiveaction})
are solved if the metric and dilaton field 
are given by \cite{lidsey,os,feinstein}
\begin{equation}
\label{curved1}
ds^2_D =H^m e^{-\frac{2}{Q(D-3)}\varphi} \tilde{f}_{\mu\nu}
dx^{\mu}dx^{\nu} +H^n e^{\frac{2}{Q} \varphi} 
dy^2
\end{equation}
and
\begin{equation}
\label{curved2}
e^{\Phi} = e^{-\frac{2}{\alpha Q}\varphi} H^{-2\alpha /\Delta} ,
\end{equation}
respectively, where
\begin{eqnarray}
\label{varphiintro}
\varphi \equiv Q B, \qquad Q \equiv \sqrt{2} \left[ 
1 +\frac{2}{\alpha^2} +\frac{1}{D-3} \right]^{1/2}
\\
\label{tildef}
\tilde{f}_{\mu\nu} =\Theta^2 f_{\mu\nu} , \qquad 
\Theta^2 \equiv e^{\frac{2}{Q(D-3)} \varphi}
\end{eqnarray}
and  $\{ \tilde{f}_{\mu\nu} , \varphi \}$ satisfy 
the $(D-1)$--dimensional Einstein field equations for a 
massless, minimally coupled scalar 
field: 
\begin{equation}
\label{Einsteineqn}
{^{(D-1)}}\tilde{R}_{\mu\nu} =\frac{1}{2} \tilde{\nabla}_{\mu} 
\varphi \tilde{\nabla}_{\nu} \varphi ,
\qquad
\tilde{\nabla}^2 \varphi =0 .
\end{equation}
A tilde denotes quantities
calculated with the conformally 
transformed world--volume metric, $\tilde{f}_{\mu\nu}$.
Since 
the 
dependence of the fields on the transverse 
coordinate is {\em identical} to that of the 
Ricci--flat solution
(\ref{isotropicbrane1})--(\ref{isotropicbrane2}), 
Eqs. (\ref{curved1})--(\ref{curved2}) 
represent a 
{\em curved} dilatonic domain wall,
where the curvature of the world--volume 
is induced by 
the non--trivial
variation of 
the modulus
field, $\varphi$. 
This 
field arises when the transverse space depends directly 
on the world--volume coordinates in an appropriate way.

We now develop curved intersecting branes 
in the type IIA supergravity theory. 
The 
Neveu--Schwarz/Neveu--Schwarz 
(NS--NS) sector of the massless theory
consists of the graviton, $\hat{g}_{AB}$, the dilaton, 
$\hat{\Phi}$, and 
the two--form potential, $\hat{A}_{(2)}$. The Ramond--Ramond (RR)
sector is comprised of one--form
and three--form 
potentials, $\hat{A}_{(1)}$
and $\hat{A}_{(3)}$ \cite{west}.
The theory admits a 5--brane supported 
by the NS--NS two-form (NS5--brane) and 
a ${\rm D}4$-- and ${\rm D}6$--brane supported 
by the RR three--form and one--forms, 
respectively. 
The massless type IIA theory 
may be derived 
from 
the  Kaluza--Klein reduction 
of eleven--dimensional supergravity (M--theory) 
on a circle, $S^1$.
The bosonic sector of 
this latter theory is 
given by Eq. (\ref{effectiveaction}), where
$(D,\alpha,q)=(11,0,4)$. 
The 
ten-- and eleven--dimensional metrics are 
related by 
$d\check{s}^2 = e^{-\hat{\Phi}/6} d
\hat{s}^2 + e^{4\hat{\Phi}/3}
( dz_{11} + \hat{A}_{(1)B}dx^B
)^2$, 
where $z_{11}$ denotes the coordinate 
parametrizing the circle and 
the conformal factors are chosen 
such that the ten--dimensional 
spacetime is the 
Einstein--frame metric \cite{9512012}. 
The corresponding field strengths are 
related by 
$\check{F}_{(4)} = \hat{F}_{(4)} + \hat{F}_{(3)} \wedge 
(dz_{11} +\hat{A}_{(1)} )$.

We  consider compactifications 
of type IIA supergravity 
to six 
dimensions
that are associated with the 
`wrapping' of a 
solitonic $p$--brane. A brane is said 
to be wrapped  
when the internal components of an antisymmetric 
tensor field are placed  onto 
a manifold, $X$, such that the form--field 
has a non--trivial flux on that space. 
The nature of the wrapping 
is
determined by the 
Betti numbers, $b_m$, of $X$, 
corresponding to 
the number of independent 
$m$--cycles in the internal manifold, 
or equivalently, to the dimensionality of the 
cohomology class, $H^m (X)$
(the set of all harmonic 
$m$--forms that are closed but not exact).
In general, wrapping a 
$p$--brane around a given $m$--cycle in $X$ leads to a 
solitonic $(p-m)$--brane. 
The simplest compactifying space is the 
$n$--dimensional torus, $T^n$, with 
Betti numbers 
$b_m=n!/[m!(n-m)!]$. The four--torus therefore 
admits 
six 
harmonic two--forms, $dy^a \wedge dy^b$ $(a,b=1,2,3,4)$, 
and this implies that there 
exist three self--dual two--forms 
and three anti--self--dual two--forms on this space: 
\begin{eqnarray}
J^{(1)}_{\pm} = dy^1 \wedge dy^4 \pm dy^2 \wedge dy^3 \nonumber \\
J^{(2)}_{\pm} = dy^2 \wedge dy^4 \pm dy^3 \wedge dy^1 \nonumber \\
J^{(3)}_{\pm} = dy^3 \wedge dy^4 \pm dy^1 \wedge dy^2  .
\end{eqnarray}

Thus, we may 
wrap the NS--NS two--form potential 
around the corresponding  
two--cycles by invoking the ansatz
$\hat{A}_{(2)} = b (x) J^{(i)}_{\pm}$,
where $b=b(x)$ is a scalar 
function that is constant over the internal 
manifold. 
When the other form fields are trivial, the truncated type IIA 
action  is given by Eq. (\ref{effectiveaction}), 
where $(D, \alpha ,q) =(10, -1 , 3)$.
Compactification of this action
on $T^4$ leads to 
ten moduli fields arising from the ${\rm SL}(4,R) \times R$ 
toroidal symmetry. For simplicity, we 
consider only the dynamics of the
breathing mode, defined by 
$\sigma \equiv -\ln \left[ \int_X \sqrt{g} \right]$, 
where $g$ is the determinant of the metric on $T^4$. 
This is equivalent to 
choosing the metric ansatz
\begin{equation}
\label{breathingmetric}
d\hat{s}^2
= e^{\sigma /2}ds^2_6 +e^{-\sigma /2}ds^2_4 (X) ,
\end{equation}
where the conformal factor 
in front of the six--dimensional line--element 
ensures that the standard Einstein--Hilbert action is recovered 
in the lower dimensions.
Assuming the ten--dimensional dilaton and the breathing mode 
to be constant 
on $T^4$ and, furthermore, that the two fields are related 
by $\hat{\Phi}=-\sigma$, 
then implies that the 
reduced
six--dimensional action can be written as 
\begin{equation}
\label{Jaction}
S= \int d^6 x \sqrt{|g|} \left[ R-\frac{1}{2}\left( \nabla x \right)^2
-\frac{1}{2} e^{-\sqrt{2} x} \left( \nabla b \right)^2 \right] ,
\end{equation}
where $x \equiv \sqrt{2} \hat{\Phi}$. 

Since the `axion' field, $b$, 
arises only through a total derivative, 
a generalized Scherk--Schwarz dimensional 
reduction to five dimensions may now be 
performed 
\cite{SchSch}. 
Compactifying on a circle 
such that 
$ds^2_6 = 
e^{-2\hat{\Phi}/3} ds^2_5 +e^{2\hat{\Phi}}dy^2_6$, 
and allowing 
the axion to 
have a linear dependence on the 
compactifying coordinate, 
$b= \Lambda y_6$, 
results in 
a five--dimensional action
of the form (\ref{effectiveaction}), where 
$(D , \alpha ,q)=(5, \sqrt{14/3} ,0)$
and $\Phi \equiv 
-\sqrt{14/3}\hat{\Phi}$. 

Thus, the theory 
admits a curved domain wall (three--brane),
where $\Delta =2$.
The five--dimensional line element is 
\begin{equation}
ds^2_5 = H^{2/3} e^{-\varphi /Q} \tilde{f}_{\mu\nu}
dx^{\mu}dx^{\nu} +H^{8/3} e^{2\varphi /Q} dy_5^2 ,
\end{equation}
where $\{ \tilde{f}_{\mu\nu} , \varphi \}$ 
satisfy the four--dimensional Einstein equations 
(\ref{Einsteineqn}) of general relativity, $Q \equiv \sqrt{27/7}$
and $H =1+m |y_5|$.
The corresponding six--dimensional metric is given by 
\begin{eqnarray}
\label{HK}
ds^2_6 = e^{-\frac{9}{7Q} \varphi} \tilde{f}_{\mu\nu} 
dx^{\mu}dx^{\nu} \nonumber \\
+H^2 \left( 
e^{\frac{12}{7Q} \varphi} dy^2_5 +e^{\frac{6}{7Q} 
\varphi} dy^2_6 \right)
\end{eqnarray}
and after oxidizing the solution back 
to ten dimensions by employing 
Eq. (\ref{breathingmetric}), we find that 
\begin{eqnarray}
\label{NS5intersect}
d\hat{s}^2 =H^{-1/2} e^{-\frac{3}{2Q} \varphi} \tilde{f}_{\mu\nu}
dx^{\mu} dx^{\nu} \nonumber \\
+H^{3/2} \left[ 
e^{\frac{3}{2Q}\varphi} dy_5^2 +e^{\frac{9}{14Q} \varphi} dy^2_6 
\right] +H^{1/2} e^{\frac{3}{14Q}\varphi} ds^2_4 \nonumber \\
e^{\hat{\Phi}} =H e^{\frac{3}{7Q} \varphi} ,
\qquad 
\hat{A}_{(2)} = \Lambda y_6 J_{(2)}  .
\end{eqnarray}
Finally, since Eq. (\ref{NS5intersect}) 
was derived within the context of 
the massless
type IIA theory,
a further oxidation to eleven dimensions 
can be made. 
We find that 
\begin{eqnarray}
\label{M5intersect}
d\check{s}^2 =  H^{-2/3} e^{-\frac{11}{7Q}\varphi} \tilde{f}_{\mu\nu} 
dx^{\mu}dx^{\nu} \nonumber \\
+H^{4/3} \left[ e^{\frac{10}{7Q}\varphi} dy^2_5
+e^{\frac{4}{7Q}\varphi} \left( dy^2_6 +dz^2_{11} \right) \right]
\nonumber \\
+ H^{1/3} e^{\frac{1}{7Q}\varphi}ds^2_4 \nonumber \\
\check{F}_{(4)} = \Lambda dy^6 \wedge J_{(2)} \wedge dz^{11}  .
\end{eqnarray}

Eq. (\ref{M5intersect}) may be interpreted 
in terms of  
M5--branes. The metric for two M5--branes 
orthogonally 
intersecting over a three--brane 
can be written as \cite{pt,ts,gkt}
\begin{eqnarray}
\label{gtmetric}
ds^2 = 
(H_1H_2)^{2/3} \left[ 
(H_1H_2)^{-1} f_{\mu\nu}dx^{\mu}dx^{\nu}
\right. \nonumber \\
\left. +H^{-1}_2 (dy^2_1 +dy^2_2 ) 
+H^{-1}_1 ( dy^2_3 +dy^2_4) \right. 
\nonumber \\
\left. +\left( dy^2_5 +dy^2_6 +dz^2_{11} \right) \right]  ,
\end{eqnarray}
where 
$H_i$ are harmonic functions over 
$(y_5,y_6,z_{11})$. When 
$H_2=1$, this metric represents a 
M5--brane with world--volume 
coordinates $(x^{\mu}, y_3, y_4)$ and transverse space spanned 
by $(y_1,y_2,y_5,y_6,z_{11})$. The brane is delocalized 
over the $(y_1,y_2)$ directions.
Similarly, when
$H_1=1$, the solution represents a M5--brane
transverse 
to $(y_3, y_4 ,y_5 , y_6, z_{11})$
and smeared over $(y_3, y_4)$.
In general, the metric 
(\ref{gtmetric}) interpolates between these two 
limits.  The transverse dependence of 
Eq. (\ref{M5intersect})
is recovered when 
$H_1=H_2$ and this 
latter solution 
may therefore 
be interpreted as two curved M5--branes 
orthogonally 
intersecting on a curved three--brane. 
Since, the harmonic function 
depends only on $y_5$, the 
M5-branes are delocalized 
over the remaining transverse dimensions. 
A similar analysis follows for the interpretation of Eq. (\ref{NS5intersect})
as the orthogonal intersection of two curved NS5--branes 
on a three--brane. 

We now consider a compactification to six
dimensions involving the RR one--form 
potential
of the type IIA theory. 
The coupling of this field to the 
ten--dimensional dilaton is given by $\alpha =3/2$.
Since this field 
arises 
only through an exterior 
derivative, 
we may 
consider 
a generalized Scherk--Schwarz compactification
on a four--dimensional manifold, $X$, 
where the closed, harmonic two--form field strength is identified 
with the cohomology class, $H^2(X)$, of 
$X$. Such a 
wrapping of the RR one--form around the four--torus 
is achieved through 
the ansatz
$\hat{F}_{(2)} = \Lambda 
J$, where $J$ is a harmonic two--form on $T^4$
and $\Lambda$ is an arbitrary constant \cite{llp}. 
For example, if 
$J=J^{(3)}_+$, 
the one--form 
is given by 
$\hat{A}_{(1)} = \Lambda \left( y^1 dy^2 + y^3 dy^4 
\right)$.
Compactifying with the metric ansatz 
(\ref{breathingmetric}), and 
equating the ten--dimensional dilaton with the breathing 
mode, $\hat{\Phi} =\sigma$, 
then 
implies that 
the six--dimensional action is given by 
Eq. (\ref{effectiveaction}), where 
$(D , \alpha, q)= (6, 3/\sqrt{2} , 0)$ and 
$\Phi  =\sqrt{2} \hat{\Phi}$. Thus, 
it follows from Eq. (\ref{mnDelta}) that 
$\Delta =2$ and Eqs. (\ref{curved1})--(\ref{Einsteineqn}) 
result in a 
domain wall solution: 
\begin{eqnarray}
ds^2_6 = H^{1/2} e^{-\frac{2}{3Q} \varphi} \tilde{f}_{\mu\nu} 
dx^{\mu}dx^{\nu} +H^{5/2} e^{\frac{2}{Q}\varphi} dy^2_6 \nonumber \\
e^{\Phi} = e^{-\frac{\sqrt{8}}{3Q} \varphi} H^{-3/\sqrt{2}}  ,
\end{eqnarray}
where $\{ \tilde{f}_{\mu\nu} , \varphi \}$ 
solve the five--dimensional Einstein 
equations (\ref{Einsteineqn}), 
$H=1+m|y_6|$ and $Q=4\sqrt{2}/3$.  
Oxidizing the solution 
back to ten dimensions then implies that
\begin{eqnarray}
\label{D6intersect}
d\hat{s}^2 = 
H^{-1/4} e^{-\varphi /Q} \tilde{f}_{\mu\nu} dx^{\mu}
dx^{\nu} \nonumber\\
+H^{7/4} e^{\frac{5}{3Q}\varphi}dy^2_6 +H^{3/4}
e^{\frac{1}{3Q} \varphi} ds^2_4  ,
\end{eqnarray}
where $e^{\hat{\Phi}} = e^{-\frac{2}{3Q} \varphi} H^{-3/2}$. 

The metric  
for two orthogonally intersecting
${\rm D}6$--branes on a ${\rm D}4$--brane is
\cite{ts,llp}
\begin{eqnarray}
ds^2 = (H_1H_2)^{7/8} 
\left[ (H_1H_2)^{-1} f_{\mu\nu} dx^{\mu}dx^{\nu} 
+dy^2_6 \right. \nonumber \\
\left. +H^{-1}_2(dy_1^2 +dy^2_2) +H^{-1}_1 (
dy^2_3 +dy^2_4) \right]  ,
\end{eqnarray}
where $H_i = 1+m_i|y_6|$. For 
example, when $H_1=1$, 
the line--element is that of a ${\rm D}6$--brane 
with world--volume and transverse coordinates $(x^{\mu}, y_1,y_2)$
and $(y_6,y_3,y_4)$.  
The two D6--branes are smoothed over the $(y_3,y_4)$ and 
$(y_1,y_2)$ directions, respectively, and when 
$H_1=H_2$, the transverse dependence 
of the metric coefficients 
reduces to that given in Eq. (\ref{D6intersect}). 
This solution therefore 
represents the orthogonal intersection 
of two curved ${\rm D}6$--branes on a 
curved four--brane, 
where the harmonic functions are identified.

To summarize thus far, we have 
found curved intersecting brane solutions by wrapping
type IIA form fields around homology cycles of the 
four--torus. 
Further
intersecting branes may now be 
generated from 
Eq. (\ref{D6intersect}) 
by employing the duality symmetries 
of string theory. We first consider the T--duality 
that maps the type IIA theory 
onto the type IIB theory, and vice--versa. We 
assume that 
all fields are independent 
of one of the world--volume coordinates $(x^5)$
and express the world--volume metric as
$ds^2_5 = e^{-\chi /\sqrt{3}} \tilde{f}^{(4)}_{\mu\nu} 
dx^{\mu}dx^{\nu} +e^{2\chi /\sqrt{3}} dx^2_5$, 
where the normalization is chosen such that 
the four--dimensional metric, $\tilde{f}_{\mu\nu}^{(4)}$, is the 
Einstein--frame metric 
and $\{ \chi , \varphi \}$ represent 
two, massless, minimally coupled scalar fields in four dimensions. 

Conformally transforming (\ref{D6intersect}) 
to 
the ten--dimensional, 
type  IIA string--frame and performing 
a T--duality in the $x^5$ direction 
results in 
a type IIB solution representing the 
intersection of two curved ${\rm D}5$--branes supported 
by the magnetic charge of the RR two--form potential, 
$B_{\mu\nu}$. 
The fields are related by $2\Phi_B =2\Phi_A 
-\ln G^{(A)}_{55}$, $G^{(B)}_{55} = 1/G^{(A)}_{55}$ and 
$B_{5\mu} =-A_{\mu}$, 
where $G$ denotes string--frame metrics
\cite{bho}. 
Applying an S--duality \cite{ht}
on the resulting 
type IIB solution then  
interchanges the RR  and NS--NS two--form
potentials  and reverses the sign of the 
dilaton, thus leading to 
a configuration consisting of two 
intersecting 
${\rm NS}5$--branes 
with a transverse dependence given by Eq. 
(\ref{NS5intersect}). 
At this level of truncation, such a solution 
also satisfies the field equations 
of the 
type IIA 
theory and it may therefore be 
oxidized
to eleven dimensions. This 
results in a further
solution of two orthogonally intersecting 
${\rm M}5$--branes: 
\begin{eqnarray}
\label{second}
d\check{s}^2= 
H^{-2/3} e^{-\frac{4}{3Q} \varphi - \frac{2}{3\sqrt{3}} \chi}
\tilde{f}^{(4)}_{\mu\nu} dx^{\mu}dx^{\nu}
\nonumber \\
+ H^{4/3} \left[ e^{\frac{4}{3Q} \varphi} \left( 
e^{-\frac{5}{3\sqrt{3}} \chi} dx^2_5 
+
e^{\frac{1}{3\sqrt{3}} \chi} dy^2_6 \right) 
\right. \nonumber \\
\left. + 
e^{\frac{4}{3\sqrt{3}}\chi} dz^2_{11} \right]
+H^{1/3} e^{\frac{1}{3\sqrt{3}}\chi} ds^2_4  .
\end{eqnarray}
Eq. (\ref{second}) is more general than Eq. 
(\ref{M5intersect}) since the world--volume is curved by two scalar 
fields. Either of these may be consistently 
set to zero. 

The wrappings around $T^4$ 
that we have considered thus far admit a
direct generalization to the 
compact, Ricci--flat, 
${\rm K}3$ manifold. 
This is 
Kummer's quartic surface in ${\rm CP}^3$
and admits 
22 
harmonic two--forms 
$(b_2=22)$.
(For a review of the 
properties of K3 surfaces, see, e.g., Ref. \cite{aspinwall}).
When discussing compactifications 
on ${\rm K}3$, it is convenient
to view it 
as an orbifold approximation to $T^4$, 
${\rm K}3 \approx T^4/{\rm Z}_2$.  
Such a compactification was considered in detail in Ref. 
\cite{lps}. 
Six of the harmonic two--forms on ${\rm K}3$
correspond to the harmonic two--forms on the four--torus. 
Thus, wrapping the NS--NS two--form potential 
of the type IIA theory around 
one of these 
two--cycles of the ${\rm K}3$ manifold results, after 
oxidation to eleven dimensions, in the 
intersecting 
brane (\ref{M5intersect}), where 
the internal 
metric, $ds^2_4$, is 
now the metric on 
${\rm K}3$. 
The intersecting ${\rm D}6$ brane 
(\ref{D6intersect}) may also be generalized to 
the ${\rm K}3$ case in a similar fashion. 

The compactification of the type IIA theory 
on ${\rm K}3$ is important in view of the 
conjectured strong/weak coupling S--duality between this theory and 
the heterotic theory compactified on $T^4$ \cite{ht,w}.
This duality implies that a curved heterotic brane may 
be derived, for example, from 
Eq. (\ref{NS5intersect}). 
The relevant transformation rules between the massless 
fields of the two theories have been summarized in Refs. 
\cite{lps,flm} 
for the case where the heterotic gauge group is broken 
to ${\rm U}(1)^{16}$. We consider the transformations relevant 
to the compactification leading to 
the type II truncated action 
(\ref{Jaction}). 
The six--dimensional string--frame metrics 
are related by 
$G_{\mu\nu}^{\rm II} = \Theta^2 
G_{\mu\nu}^{\rm het}$, 
where $\Theta^2 \equiv 
e^{-2\psi_{\rm het}}$, 
and the six--dimensional dilatons are given by 
$\psi_{\rm II} =
-\psi_{\rm het}$.
In the above type II compactification, 
we have only considered the 
breathing mode of the ${\rm K}3$ manifold 
and this is equivalent to 
assuming that all four radii of the orbifold $T^4/{\rm Z}_2$ 
are equal, i.e., $\hat{G}_{aa}^{\rm II} =
\hat{G}_{bb}^{\rm II}$ $(a,b=1,2,3,4)$. 
This places a restriction on the 
toric radii in the corresponding compactification 
of the heterotic theory. Specifically, 
in the ten--dimensional string frame, three of the internal 
dimensions are static, $\hat{G}_{ii}^{\rm het} =1$,
and the radius of the fourth is given by 
$\hat{G}_{44}^{\rm het} = (\hat{G}^{\rm II}_{aa})^2$. 
Finally, the scalar axion field, $b$,  
arising from the wrapping of the NS--NS 
two--form potential around the ${\rm K}3$ 
two--cycle, is related to 
one of the sixteen ${\rm U}(1)$ potentials, 
$\hat{A}^{\rm het}_{\mu}$, such that 
$b=A_4 /\sqrt{2}$, where the scalar 
field, $A_4$, arises 
from 
compactification of 
the ${\rm U}(1)$ gauge field 
on the circle 
parametrized by $y^4$, i.e., 
$\hat{A}^{\rm het} = A_4 dy^4$. 

These type II/heterotic 
correspondences 
may therefore 
be employed to derive the 
curved heterotic brane  
that is S--dual 
to the 
type IIA, six--dimensional metric (\ref{HK}). 
Oxidizing the resulting solution 
to ten dimensions then yields the heterotic 
solution: 
\begin{eqnarray}
\label{6brane}
d\hat{s}^2_{\rm het}= H^{-1/4} \left[ e^{-\frac{39}{28Q} \varphi} 
\tilde{f}_{\mu\nu}^{(4)} dx^{\mu}dx^{\nu} 
\right. \nonumber \\
\left. + e^{-\frac{3}{28Q}\varphi} 
\left( dy^2_1 +dy^2_2
+dy^2_3 \right) \right] \nonumber \\
+H^{7/4}
e^{\frac{3}{4Q} \varphi} \left( dy^2_4 +
e^{\frac{6}{7Q} \varphi} dy^2_5 +dy^2_6 \right)
\nonumber \\
e^{\hat{\Phi}_{\rm het}} 
=
H^{1/2} e^{\frac{3}{14Q}\varphi} ,
\end{eqnarray}
where $Q= \sqrt{27/7}$, 
$\hat{F}^{\rm het}_{(2)} = \sqrt{2} \Lambda dy^6 \wedge dy^4$
and the metric is expressed in the Einstein frame. 
Eq. (\ref{6brane}) 
represents a curved six--brane, 
where  
three of the transverse 
dimensions of the type IIA solution 
(\ref{NS5intersect}) 
have become world--volume dimensions 
in the heterotic solution.
The S--duality between the ${\rm SO}(32)$ heterotic 
and type I theories may also be invoked to 
derive the corresponding type I brane \cite{w,hetI}. 
The vacuum limit $(\varphi =0)$ 
of 
Eq. (\ref{6brane}) is the 
heterotic/type I six--brane 
found in Ref. \cite{bbhp} 
that is T--dual to the bound state 
of an  anti five--brane and a 
Kaluza--Klein monopole. Such a solution 
arises as a special case in  
the domain--wall/quantum--field-theory 
correspondence \cite{dwqft,bbhp}.

The six--brane (\ref{6brane}) 
is also relevant to the recently introduced 
compactification ansatz referred to as
`braneworld Kaluza--Klein reduction' 
\cite{bb}. In this scheme, the world--volume of 
a co--dimension one brane arising as a solution 
of a gauged supergravity theory 
is determined by 
an ungauged supergravity 
theory with half the supersymmetry. 
For example, 
the  
massive type IIA supergravity theory 
of Romans \cite{romans}
admits a ${\rm D}8$--brane, 
where the curvature of the world--volume is 
determined by a solution to nine--dimensional, ungauged 
$N=1$ supergravity. 
Under appropriate conditions, this
latter theory may be derived by 
a Kaluza--Klein compactification on a circle 
of the (truncated) ten--dimensional 
type I
theory. It follows, therefore, 
that the   
dimensional
reduction of 
the six--brane (\ref{6brane})
along 
a world--volume coordinate 
results 
in a 
five--brane 
of 
$D=9$, $N=1$ supergravity. 
Consequently, following the prescription outlined in 
Ref. \cite{bb}, such a brane may be embedded 
within the ${\rm D}8$ solution 
of the massive type IIA theory.
The resulting configuration 
corresponds to the 
intersection of an NS5--brane and a D6--brane with a D8--brane.  

More general solutions 
to those presented may be found 
by noting 
that
the scalar fields 
$\{ x, b \}$ in action (\ref{Jaction})
parametrize the ${\rm SL}(2,R)/{\rm U}(1)$ 
coset. The action
is therefore invariant 
under a global ${\rm SL}(2,R)$ symmetry transformation,   
where 
the complex scalar field, $\kappa \equiv 
\lambda b +ie^{\lambda x}$ $(\lambda \equiv 1/\sqrt{2})$,
undergoes a fractional linear transformation
$\bar{\kappa} = (A \kappa +B)/(C \kappa +D)$
for 
$AD-BC=1$,
and the Einstein--frame metric transforms as a singlet. 
Given a solution $(b ,x)$ to the field equations
derived from Eq. (\ref{Jaction}), the 
${\rm SL}(2,R)$ transformation 
may be employed to generate a  
class of solution where both fields have a non--trivial 
dependence on the transverse and world--volume 
coordinates. Moreover, 
a generalized Scherk--Schwarz compactification 
of action (\ref{Jaction}) may also 
be performed, where 
the dependence of the fields on the compactifying 
coordinates is determined by a local 
${\rm SL}(2,R)$ transformation, thereby extending the 
linear ansatz we invoked for the axion field
\cite{general}. 

A related six--dimensional 
${\rm SL}(2,R)/{\rm U}(1)$ model 
is derivable by compactifying 
eight--dimensional, vacuum Einstein gravity on  
a non--dynamical two--torus, 
$ds^2_8 =ds^2_6+e^{-\Phi}dy^2_6
+e^{\Phi} (dy_5+\sigma dy_6)^2$; the 
$\{ \Phi , \sigma \}$ fields 
parametrize the 
coset manifold, $ds^2 = d\Phi^2 + e^{2\Phi}
d\sigma^2$, and therefore support 
a 3-brane after Scherk--Schwarz 
compactification to five dimensions.
This is interesting because a mapping 
between $D=8$ vacuum Einstein gravity 
with two commuting spacelike isometries and 
$D=11$ supergravity was recently 
established by means of a non--local classical duality
\cite{cgs}. 
Thus, a given solution to 
one theory acts as a seed for generating new 
solutions in the other, and vice--versa. 
Indeed, such a correspondence 
has been employed to generate 
intersecting ${\rm M}5$--branes \cite{cgs}. 
The results of the present work imply that 
analogous curved models may 
also be 
found by this procedure. 
It is also worth remarking 
that branes intersecting at angles can be 
derived by
applying successive T--S--T duality transformations 
on orthogonally intersecting configurations 
\cite{gauntlett,f}. 
It would be interesting to explore 
such a procedure 
to derive tilted curved branes. 

Finally, we conclude by 
discussing 
some of the 
cosmological implications of the solutions 
we have derived. 
Considerable interest has 
been generated recently by the 
proposal 
that 
our observable, 
four--dimensional universe 
corresponds to a domain wall or 
$p$--brane embedded in a higher--dimensional 
space \cite{braneref,hw,lukas,RS}. 
A natural generalization 
of the simplest braneworld scenario is to 
view 
our universe as the intersection of two or more 
higher--dimensional branes \cite{intersecting}.
The solutions we have found 
can be interpreted cosmologically 
when the scalar field 
in Eq. (\ref{Einsteineqn}) 
is time--dependent. 
Since the world--volume, 
$\tilde{f}_{\mu\nu}$, 
is arbitrary, 
a wide class of 
spatially anisotropic and inhomogeneous 
cosmologies may be considered that 
generalize the standard Friedmann--Robertson--Walker 
(FRW) models.  
This is important since deviations from spatial 
isotropy are expected to have been 
significant in the very early universe. 

To be specific, 
curved 
braneworlds may be found directly 
once a 
solution to 
Eq. (\ref{Einsteineqn}) has been 
given. Eq. 
(\ref{Einsteineqn}) represents 
Einstein's equations sourced 
by a massless, minimally coupled scalar field
and solutions to this latter theory are 
known \cite{exact,lazkoz,lwc}. 
In particular, 
homogeneous and inhomogeneous 
models containing one or more 
massless scalar fields 
were recently reviewed \cite{lwc}
within the context of string--inspired models
such as the pre--big bang inflationary cosmology \cite{pbb}.
In this latter scenario, inflation can be interpreted in the 
Einstein--frame as the collapse of a scalar field
dominated universe, 
where the dynamics is determined by Eq. (\ref{Einsteineqn})
\cite{bh}.
Thus, our solutions provide a framework for 
considering pre--big bang inflation in a braneworld setting. 

One of the 
simplest cosmological models 
is represented by the spatially 
homogeneous and anisotropic Bianchi type I metric. 
When the world--volume 
has this form, it can 
be shown that 
the eleven--dimensional 
metrics that 
we have derived 
correspond to {\em vacuum} solutions of Einstein gravity
in the limit where
the harmonic function $H=1$. 
Since the spatial hypersurfaces 
are Ricci--flat, 
these metrics represent higher--dimensional generalizations of the 
four--dimensional 
Kasner solution \cite{kasner} and 
it is known that 
for these models, 
inflation is possible 
over a wide 
region of parameter space \cite{lwc}. 
The accelerated expansion 
of a subset of the spatial dimensions 
is driven by the collapse of the 
remaining dimensions \cite{levin}. 
It would be interesting to investigate 
inflation of this type within 
the intersecting braneworld context, 
although a 
detailed analysis 
is beyond the scope of the present paper. 

An important question in the braneworld scenario 
is whether gravity 
can be localized on 
the domain 
wall. If the 
transverse dimension is 
infinite, a 
necessary condition 
for gravity to be confined 
on a brane of the 
type given 
in Eq. (\ref{isotropicbrane1})
is that 
$-2(D-1)/(D-2) \le   \Delta \le -2$ \cite{clp}. 
In the models considered above, 
$\Delta =2$. However, 
in this case, gravity can arise on the world--volume
of the brane if the extra 
coordinate is compact. 
For example, 
if the coordinate is restricted to the interval, 
$S^1/{\rm Z}_2$, the domain wall may 
be located on the orbifold fixed points, 
as in the Ho\v{r}ava--Witten theory \cite{hw,lukas}. 
(This model corresponds to 
$\alpha = -2$ in Eq. (\ref{effectiveaction})). 

A
significant consequence of viewing our 
observable universe as a co--dimension one brane embedded in a 
five--dimensional `bulk' space is that the effective 
four--dimensional gravitational field equations  
include extra terms. 
These `tidal effects' are 
parametrized by the Weyl tensor of the 
higher--dimensional metric 
and do not depend 
specifically on
the energy--momentum 
of matter that is confined to the brane \cite{weyl}. 
Hence, the geometry of the bulk 
can significantly influence the 
lower--dimensional brane dynamics
and  
in general this implies 
that
the cosmological expansion of the brane 
can not be determined 
unless the form of the higher--dimensional metric
is known. 
In this paper, we have found exact bulk 
solutions to the type II string theory and 
M--theory field equations and these solutions 
therefore provide a class of models where 
the cosmological dynamics of the 
brane can be determined. 

Recently, 
Feinstein, Kunze and 
V\'azquez--Mozo considered 
a related class of five--dimensional
domain wall models supported by an exponential 
potential of the form given in Eq.
(\ref{effectiveaction}) \cite{feinstein}. 
These authors included a 
matter source confined to the brane with 
a Lagrangian 
coupled to the scalar field via a Liouville term. 
In this case, a self--tuning mechanism arises 
between the matter and the brane tension 
that causes 
the effective cosmological constant on the 
brane to vanish. In principle, 
a similar analysis may be performed 
for
the intersecting brane configurations
derived above by introducing an appropriate 
matter source.

\acknowledgements

The author is supported by the Royal Society.

\end{document}